\documentclass{pasj01}

\RequirePackage{mathpazo}
\RequirePackage[T1]{fontenc}

\def\commenta{$^*$}
\def\commentb{$^\dagger$}
\def\commentc{$^\ddagger$}

\newcounter{author}
\def\authorcount#1#2{\refstepcounter{author}\label{#1}
                     \altaffiltext{\ref{#1}}{#2}}

\begin{document}
\SetRunningHead{T. Kato et al.}{SU UMa-Type Dwarf Nova with Long Precursor Outburst}

\Received{201X/XX/XX}
\Accepted{201X/XX/XX}

\title{CS~Indi: SU UMa-Type Dwarf Nova with Long Precursor Outburst}

\author{Taichi~\textsc{Kato},\altaffilmark{\ref{affil:Kyoto}*}
        Franz-Josef~\textsc{Hambsch},\altaffilmark{\ref{affil:GEOS}}$^,$\altaffilmark{\ref{affil:BAV}}$^,$\altaffilmark{\ref{affil:Hambsch}}
        Berto~\textsc{Monard},\altaffilmark{\ref{affil:Monard}}$^,$\altaffilmark
{\ref{affil:Monard2}}
        Peter~\textsc{Nelson},\altaffilmark{\ref{affil:Nelson}}
        Rod~\textsc{Stubbings},\altaffilmark{\ref{affil:Stubbings}}
        Peter~\textsc{Starr},\altaffilmark{\ref{affil:Starr}}
}

\authorcount{affil:Kyoto}{
     Department of Astronomy, Kyoto University, Kyoto 606-8502, Japan}
\email{$^*$tkato@kusastro.kyoto-u.ac.jp}

\authorcount{affil:GEOS}{
     Groupe Europ\'een d'Observations Stellaires (GEOS),
     23 Parc de Levesville, 28300 Bailleau l'Ev\^eque, France}

\authorcount{affil:BAV}{
     Bundesdeutsche Arbeitsgemeinschaft f\"ur Ver\"anderliche Sterne
     (BAV), Munsterdamm 90, 12169 Berlin, Germany}

\authorcount{affil:Hambsch}{
     Vereniging Voor Sterrenkunde (VVS), Oude Bleken 12, 2400 Mol, Belgium}

\authorcount{affil:Monard}{
     Bronberg Observatory, Center for Backyard Astrophysics Pretoria,
     PO Box 11426, Tiegerpoort 0056, South Africa}

\authorcount{affil:Monard2}{
     Kleinkaroo Observatory, Center for Backyard Astrophysics Kleinkaroo,
     Sint Helena 1B, PO Box 281, Calitzdorp 6660, South Africa}

\authorcount{affil:Nelson}{
     1105 Hazeldean Rd, Ellinbank 3820, Australia}

\authorcount{affil:Stubbings}{
     Tetoora Observatory, 2643 Warragul-Korumburra Road, Tetoora Road,
     Victoria 3821, Australia}

\authorcount{affil:Starr}{
     Warrumbungle Observatory, Tenby, 841 Timor Rd,
     Coonabarabran NSW 2357, Australia}



\maketitle

\begin{abstract}
We observed the 2018 November outburst of CS~Ind
and confirmed that it was a genuine superoutburst
with a very long [0.12471(1)~d in average]
superhump period.
The superoutburst was preceded by a long precursor,
which was recorded first time in SU UMa-type dwarf novae.
We interpret that the combination of a sufficient
amount of mass in the disk before the ignition of
the outburst and the slow development of
tidal instability near the borderline of
the 3:1 resonance caused a cooling front to start
before the full development of tidal instability.
This finding provides another
support to the recent interpretation of
slow development of the tidal instability
causing various phenomena similar to WZ Sge-type
dwarf novae in SU UMa-type dwarf novae with
very long orbital periods.
\end{abstract}

\section{Introduction}

   Cataclysmic variables (CVs) are composed of a white dwarf
and a mass-transferring secondary, and the transferred
matter forms an accretion disk around the white dwarf.
Thermal instability in the accretion disk causes outbursts
in some cataclysmic variables and these objects are called
dwarf novae.  SU UMa-type dwarf novae are a class of
dwarf novae which show long, bright outbursts (superoutbursts)
in addition to normal outbursts,
and these superoutbursts are believed to be caused by
the tidal instability when the disk radius expands to
the 3:1 resonance during an outburst \citep{osa89suuma}.
During superoutbursts, variations with periods slightly
longer than the orbital period ($P_{\rm orb}$)
are observed and they are called superhumps.
Dwarf novae which show only normal outbursts are called
SS Cyg-type dwarf novae.
[for general information of CVs,
dwarf novae, SU UMa-type dwarf novae and superhumps,
see e.g. \citet{war95book}].

   As CVs evolve, the secondary star loses the mass
and the orbital period shortens until the secondary becomes
degenerate.  During this evolution, the mass ratio ($q=M_2/M_1$)
becomes smaller.  In large $q$ systems, the radius of
the 3:1 resonance cannot be inside the tidal truncation radius
or even inside the Roche lobe in the extreme case,
and there is an upper limit of $q$ for a system to be
an SU UMa-type dwarf nova.  This limit is suggested to be
around 0.24 based on modern 3-D numerical simulation
\citep{smi07SH} and it can be larger (0.33) under condition
of reduced mass-transfer \citep{mur00SHintermediateq}.
Observations have shown that orbital periods of
SU UMa-type dwarf novae are almost
exclusively below 0.11~d (cf. \cite{Pdot9}).
The single traditional exception was TU~Men
(\cite{sto81tumen1}; \cite{sto84tumen}),
whose orbital and superhump periods
are 0.1172~d and 0.1257~d, respectively
\citep{men95tumen}.  In 2016, this record was broken
by the discovery of OT J002656.6$+$284933
(CSS101212:002657$+$284933) with a superhump period
of 0.13225(1)~d \citep{kat17j0026}, although a few more objects
had been suggested to be SU UMa-type dwarf novae with
longer periods, but with poorer statistics
[\citet{mro13OGLEDN2}; see discussion in Note added in proof in
\citet{kat17j0026}].  OT J002656.6$+$284933 was shown
to have a smaller $q$ than expected from the superhump
period and it was suggested that the condition of
the 3:1 resonance is difficult to meet in long-$P_{\rm orb}$
systems \citep{kat17j0026}.

   We report on a discovery of the SU UMa-type dwarf nova
CS~Ind with a long superhump period ($\sim$0.125~d).
This object not only showed ordinary evolution of
superhumps as in other SU UMa-type dwarf novae but also
showed a long-lasting precursor outburst, which has been
recorded for the first time in SU UMa-type dwarf novae.

\section{Observation and Analysis}

   CS~Ind (=NSV 13983) has been known as a dwarf nova.
Based on an approximate measurement of
long $P_{\rm orb}$ of 0.11~d
\citep{con08nsv13983} and the historical light curve
in the ASAS-3 data \citep{ASAS3}, this object had been
regarded as an SS Cyg-type dwarf nova.
Upon visual detection of a precursor outburst by R.~Stubbings
and a subsequent bright outburst in 2018 November
(vsnet-alert 22783)\footnote{
   VSNET alert message can be accessed at
   $<$http://ooruri.kusastro.kyoto-u.ac.jp/\\
pipermail/vsnet-alert/$>$.
}, this object was suspected to be
an SU UMa-type dwarf nova and the nature was finally
confirmed by the detection of superhumps
(vsnet-alert 22796).
   We observed this object by the VSNET
Collaboration \citep{VSNET}.  The observers used 35--44cm
telescopes.  All observers used unfiltered CCD cameras.
They used aperture photometry and extracted magnitudes
relative to comparison stars whose constancy has been 
confirmed by comparison with check stars.  The remaining
small zero-point differences between observers
were corrected by adding constants to minimize the squared
sum of adjacent observations in the combined light curve.
The analysis of superhumps was performed
in the same way as described in \citet{Pdot} and \citet{Pdot6}.
We mainly used R software\footnote{
   The R Foundation for Statistical Computing:\\
   $<$http://cran.r-project.org/$>$.
} for data analysis.
We used locally-weighted polynomial regression 
(LOWESS: \cite{LOWESS}) for de-trending the data
before analysis of superhumps.
The times of superhumps maxima were determined by
the template fitting method as described in \citet{Pdot}.
The times of all observations are expressed in 
barycentric Julian days (BJD).

\section{Results and Discussion}

\subsection{Course of Outburst and Long-Term Behavior}

\begin{figure}
  \begin{center}
    \FigureFile(85mm,70mm){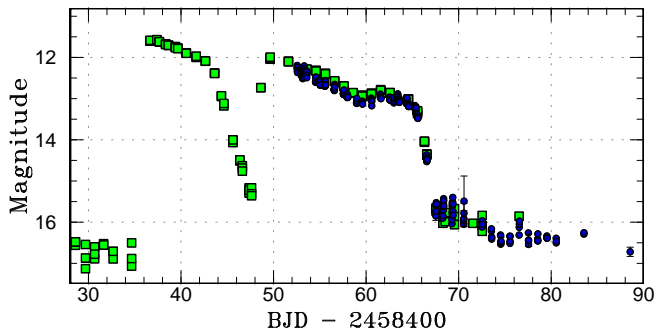}
  \end{center}
  \caption{Light curve of CS~Ind during the 2018 November
  outburst.  Filled circles and filled squares represent
  our CCD photometry and ASAS-SN data, respectively.
  A long-lasting precursor outburst followed by a dip
  and a superoutburst is clearly seen.
  }
  \label{fig:csindlc}
\end{figure}

   Although it was not known at the time of the detection
of a precursor outburst by R.~Stubbings, the entire course of
the outburst was recorded by
All-Sky Automated Survey for Supernovae (ASAS-SN)
Sky Patrol (\cite{ASASSN}; \cite{koc17ASASSNLC})
(cf. vsnet-alert 23339).
The course of the outburst was a typical one
(cf. \cite{Pdot}) for an SU UMa-type dwarf nova
except the initial long-lasting precursor, which will be
discussed later (figure \ref{fig:csindlc}).

\begin{figure}
  \begin{center}
    \FigureFile(85mm,110mm){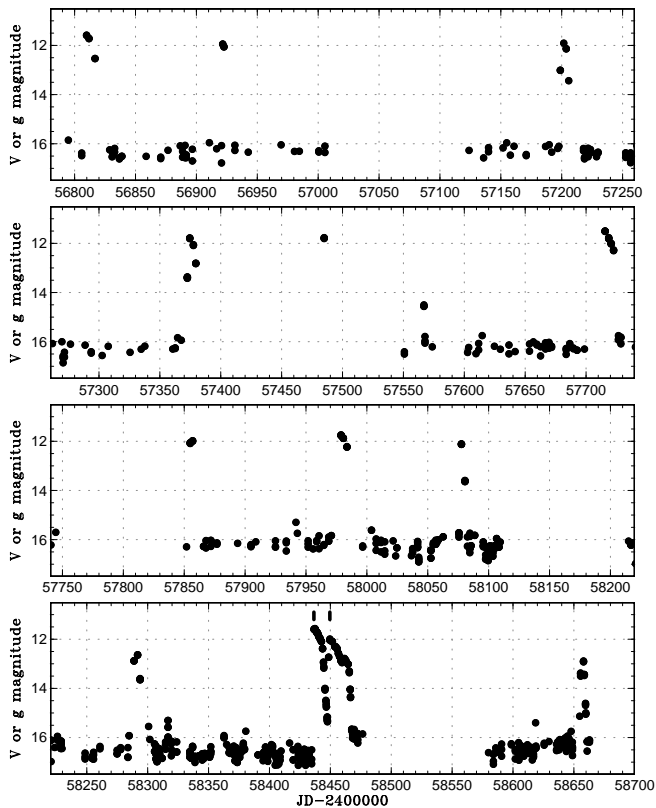}
  \end{center}
  \caption{Light curve of CS~Ind based on the ASAS-SN data.
  The vertical ticks in the bottom panels represents
  the precursor outburst and superoutburst.}
  \label{fig:csindlonglc}
\end{figure}

   The long-term behavior based on the ASAS-SN data
is shown in figure \ref{fig:csindlonglc}.
Most outbursts lasted shorter than 10~d and the general
behavior resembled that of an SS Cyg-type dwarf nova.
Only the 2018 November outburst (figure \ref{fig:csindlc}
and in the final panel of figure \ref{fig:csindlonglc})
was peculiar with a long precursor,
a subsequent dip and a long outburst.
The entire duration of this event was at least 30~d.
No similar outburst was recorded in the ASAS-3 data
between 2001 and 2009.  Visual observations by R.~Stubbings
between 2010 and 2014 did not record such an outburst.

\subsection{Superhumps}

\begin{figure}
  \begin{center}
    \FigureFile(85mm,110mm){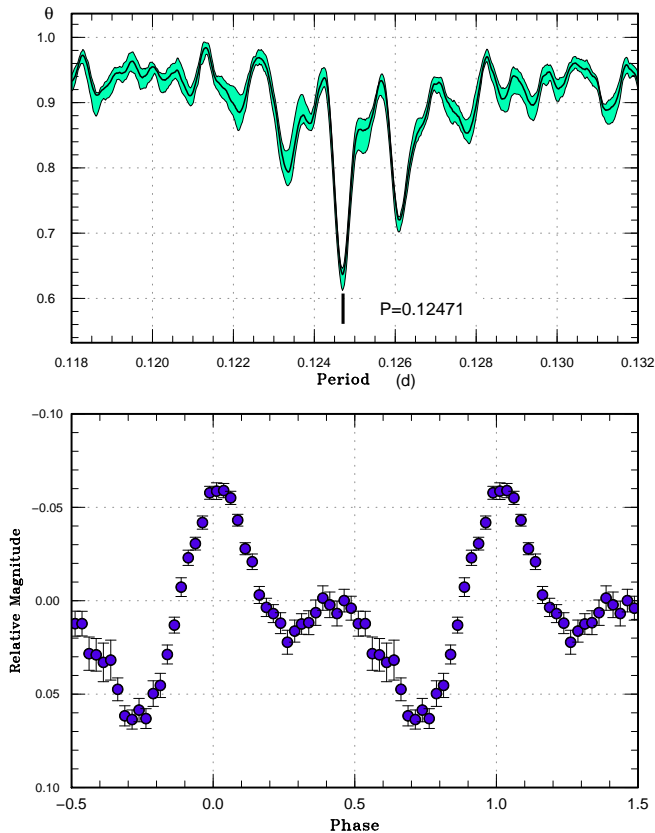}
  \end{center}
  \caption{Superhumps in CS~Ind during the superoutburst plateau.
     (Upper): PDM analysis.
     (Lower): Phase-averaged profile.}
  \label{fig:csindpdm}
\end{figure}

   We first demonstrate the presence of superhumps.
Figure \ref{fig:csindpdm} shows a phase dispersion minimization
(PDM, \cite{PDM}) analysis of the plateau phase
(BJD 2458452--2458466) of the outburst.
Superhumps were already present at the start of our campaign.
The mean superhump period during the outburst plateau
was 0.12471(1)~d [the error was estimated by the methods
of \citet{fer89error} and \citet{Pdot2}].

   The times of superhump maxima are listed in table
\ref{tab:csindoc}.  Due to the long superhump period
and the limited duration of nightly observations, the maxima
of superhumps at later epochs fell in the observing runs
and the times of maxima were not determined.
The superhump variation, however, was well present
and there was no doubt that superhumps were present
throughout the entire superoutburst plateau.
The light curve of the superoutburst showed slight
brightening after BJD 2458460.  This feature is usually
associated with the stage transition of superhumps
(called stages B and C, see \cite{Pdot}).  The mean
superhump periods before and after this were
0.12638(3)~d and 0.12488(3)~d, respectively.
The period decreased by 1.2\%, which is a typical
value for stage B-C transition \citep{Pdot}.
We conclude that the object showed a pattern
of stage B-C transition typical for an SU UMa-type
dwarf nova.  The object was already stage B at the start
of observation and stage A was not observed.

\begin{table}
\caption{Superhump maxima of CS~Ind}\label{tab:csindoc}
\begin{center}
\begin{tabular}{rp{55pt}p{40pt}r@{.}lr}
\hline
\multicolumn{1}{c}{$E$} & \multicolumn{1}{c}{max\commenta} & \multicolumn{1}{c}{error} & \multicolumn{2}{c}{$O-C$\commentb} & \multicolumn{1}{c}{$N$\commentc} \\
\hline
0 & 58452.5577 & 0.0008 & 0&0011 & 39 \\
3 & 58452.9214 & 0.0043 & $-$0&0157 & 33 \\
4 & 58453.0637 & 0.0006 & $-$0&0002 & 140 \\
5 & 58453.2016 & 0.0018 & 0&0109 & 43 \\
6 & 58453.3163 & 0.0004 & $-$0&0013 & 256 \\
7 & 58453.4521 & 0.0014 & 0&0077 & 76 \\
8 & 58453.5675 & 0.0005 & $-$0&0037 & 25 \\
16 & 58454.5915 & 0.0030 & 0&0057 & 23 \\
24 & 58455.5968 & 0.0015 & $-$0&0036 & 39 \\
32 & 58456.6142 & 0.0024 & $-$0&0009 & 27 \\
\hline
  \multicolumn{6}{l}{\commenta BJD$-$2400000.} \\
  \multicolumn{6}{l}{\commentb Against max $= 2458452.5566 + 0.126827 E$.} \\
  \multicolumn{6}{l}{\commentc Number of points used to determine the maximum.} \\
\end{tabular}
\end{center}
\end{table}

   On two nights after the superoutburst (between
BJD 2458468 and 2458470), when dense observations were
available, a period of 0.1242(1)~d was detected.
This period probably represent that of late-stage
superhumps.  No superhump signal was confirmed after
these observations.

\subsection{Long Precursor Outburst}

   The most striking feature of this outburst was
the long precursor preceding the well-confirmed superoutburst.\footnote{
   We call this phenomenon ``precursor'' since it occurred
   before the superoutburst as defined by the presence
   of well-defined superhumps.
}
Such behavior was never recorded in SU UMa-type
dwarf novae.  Precursors in SU UMa-type dwarf novae
are considered to occur when the disk reaches the 3:1 resonance
[see figure 4 in \citet{osa03DNoutburst} and \citet{osa05DImodel}].
A cooling wave immediately starts (as in normal outbursts)
and ignition of the tidal instability causes brightening to
the superoutburst.  In this picture, there is no room for
a long precursor.

   The entire light curve of the 2018 November outburst
may look like a ``type-E'' superoutburst of WZ Sge-type
dwarf novae [see e.g. figure 9 in \citet{kat15wzsge}
or figures 3 and 7 in \citet{kim18asassn16dt16hg}].
WZ Sge-type dwarf novae are an extreme extension of
SU UMa-type dwarf novae and their $q$ values are small
enough to enable the 2:1 resonance \citep{osa02wzsgehump}
and the ``type-E'' superoutburst is the most extreme
type of outburst when $q$ is extremely low \citep{kat15wzsge}.
During ``type-E'' superoutburst, the initial outburst
is interpreted to be dominated by the 2:1 resonance,
which suppresses the 3:1 resonance \citep{lub91SHa}.
Once the phase of the 2:1 resonance ends, the 3:1 resonance
starts to develop.  In very low-$q$ systems, the growth
time of the 3:1 resonance is very slow \citep{lub91SHa}
enabling a dip or an even longer faint state before
the 3:1 resonance brings the disk back to the outburst
state again \citep{kat13j1222}.  This picture is less
likely applicable to CS~Ind since the 2:1 resonance
is impossible to achieve in such a long-$P_{\rm orb}$
(i.e. high $q$) system.

   The most likely explanation of the present event
is that the sufficient amount of mass was stored
in the disk before the precursor, and when the outburst
started, the radius of the expanded disk stayed
around the tidal truncation radius for a long time
as in figure 5 in \citet{osa05DImodel}.
In ordinary SU UMa-type dwarf novae, tidal instability
develops quickly enough before a cooling wave starts
and there is no dip during the superoutburst.
In CS~Ind, it is most likely that the development of
the 3:1 resonance was slow enough, as in the case of
low-$q$ WZ Sge-type dwarf novae, which enabled a cooling
wave to start.

   Slow growth of the 3:1 resonance has been suggested
to interpret the unusual behavior (simulating WZ Sge-type
dwarf novae) in long-$P_{\rm orb}$ SU UMa-type
dwarf novae close to the stability border of
the 3:1 resonance.  Such behavior includes
long duration of growing phase of superhumps
(subsection 4.7 in \cite{Pdot6}) and
post-superoutburst rebrightenings \citep{kat16v1006cyg}.

   The present finding of CS~Ind provides additional support to
the slow growth of the 3:1 resonance when the 3:1 resonance
is difficult to excite near the stability border.
It has been well known that the historical exceptional
object TU~Men has three types of outbursts
(normal outbursts, long normal outbursts and superoutbursts),
and long normal outbursts are considered to
be outbursts starting with sufficient amount of mass
in the disk but failed to excite the 3:1 resonance.
Recently, two long-$P_{\rm orb}$ systems have been
confirmed to show these three types of outbursts:
NY~Ser (\cite{pav14nyser}; \cite{kat19nyser})
V1006~Cyg \citep{kat16v1006cyg}.

   Looking at the long-term light curve in 
figure \ref{fig:csindlonglc},
CS~Ind also showed both short and long outbursts.
The long outbursts can be thus regarded as long normal
outbursts (pending observations confirming the lack
of superhumps) and this object is expected to join
this small group of SU UMa-type dwarf novae.

   In such systems, it is apparent that the tidal
instability is very difficult to excite considering
the low number of genuine superoutbursts: three times
in TU~Men in the past 56 years, once in V1006~Cyg
in the past 13 years, once in CS~Ind in the past
19 years.  The case in NY Ser is more complex
which showed standstills and two superoutbursts
in 2018 \citep{kat19nyser}.  Before them, only one
superoutburst was documented in 1996 \citep{nog98nyser}.
Further systematic observations of these extreme
objects will clarify the behavior of the disk
around the stability limit of the tidal instability.

\subsection{Candidate Object}

   While studying the ASAS-SN light curves of dwarf novae,
one of the authors (T.K.) noticed the presence of double
outburst in V544~Her in 2018 August similar to CS~Ind.
Since such outbursts are hardly met in SS Cyg-type
dwarf novae, this object would be a candidate SU UMa-type
dwarf nova close to the stability limit of
the tidal instability.  Previous observations of two
long outbursts of this object did not show superhumps
(VSNET Collaboration, unpublished).  Future observations
of (probably) rare double outburst or the measurement
of the orbital parameters would clarify this possibility.

\begin{figure}
  \begin{center}
    \FigureFile(85mm,110mm){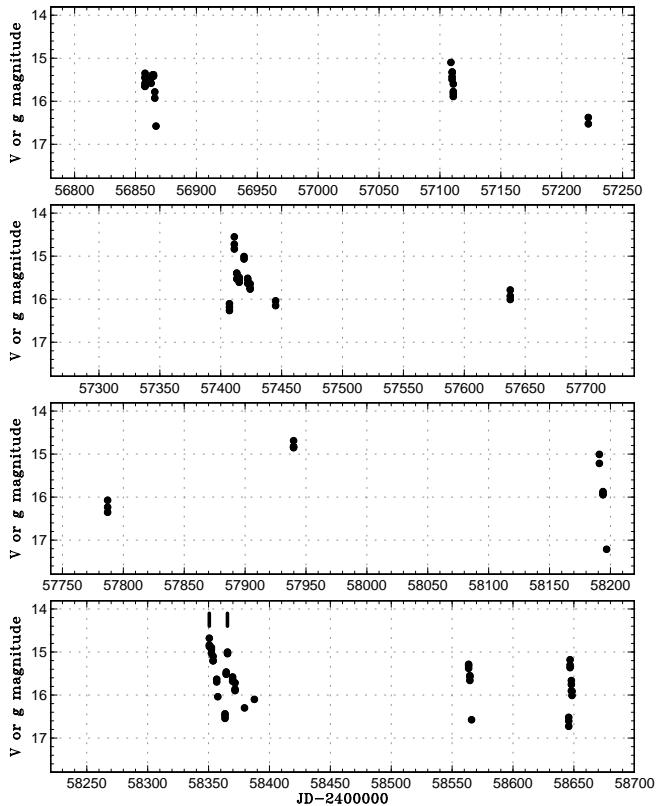}
  \end{center}
  \caption{Light curve of V544 Her based on the ASAS-SN data.
  The vertical ticks in the bottom panels represents
  double outbursts.}
  \label{fig:v544herlonglc}
\end{figure}

\section*{Acknowledgments}

   The authors express thanks to Y. Wakamatsu, who
helped processing the data reported to VSNET.

\section*{Supporting information}

   The raw observation data can be found in the online version
of this article: indcs.bjd. \\
Supplementary data is available at PASJ Journal online.

\end{document}